\documentclass{IEEE_lsens}
\usepackage[utf8]{inputenc}
\usepackage{textcomp}
\usepackage[noadjust]{cite}
\ifCLASSINFOpdf
\else

\fi

\usepackage[T1]{fontenc} 
\usepackage{amsmath}
\usepackage{amssymb}
\usepackage[cmintegrals]{newtxmath}
\usepackage{bm}
\usepackage{array}
\usepackage{url}
\usepackage{graphicx}
\usepackage{rotating}
\usepackage{float}
\usepackage{subcaption}
\usepackage{blindtext}
\usepackage{multirow}
\usepackage{graphics}

\renewcommand{\exp}[1]{{\sf exp}\left (#1\right )}

\ifCLASSINFOpdf

\else

\fi

\providecommand{\hypersetup}[1]{\relax}

\begin{document}

\IEEELSENSarticlesubject{}

\title{Impulsive Noise Detection for Intelligibility and Quality Improvement of Speech Enhancement Methods Applied in Time-Domain}

\author{\IEEEauthorblockN{C. Medina and R. Coelho\IEEEauthorieeemembermark{1}}
\IEEEauthorblockA{Laboratory of Acoustic Signal Processing, Military
Institute of Engineering, Rio de Janeiro, 22290-270, Brazil\\
\IEEEauthorieeemembermark{1}Senior Member, IEEE}
\thanks{Corresponding author: R. Coelho (e-mail: coelho@ime.eb.br)}
}

\IEEEtitleabstractindextext{%
\begin{abstract}
This letter introduces a novel speech enhancement method in the Hilbert-Huang Transform domain to mitigate the effects of
acoustic impulsive noises. 
The estimation and selection of noise components is based on the impulsiveness index of decomposition modes.
Speech enhancement experiments are conducted considering five acoustic noises with different impulsiveness
index and non-stationarity degrees under various signal-to-noise ratios.
Three speech enhancement algorithms are adopted as baseline in the evaluation analysis considering spectral and time domains.
The proposed solution achieves the best results in terms of objective quality measures and
similar speech intelligibility rates to the competitive methods.
\end{abstract}

}

\maketitle

\section{Introduction}
\label{sec:intro}

\IEEEPARstart{I}{mpulsive} background noisy condition may cause severe impact on the accuracy of 
acoustic classification systems and applications. 
Impulsive noises (slamming doors, industrial machinery, falling objects) are encountered in real environments.
They are commonly characterized by almost instantaneous sharp sounds with high acoustic energy and wide spectral bandwidth. 
Impulsive sample sequences are generally defined in the literature by heavy-tail distributions 
tailored by its impulsiveness degree. 
Due to this impulsive nature, a key element of the research area includes accurate estimation of noise components especially
from real acoustic noisy signals.

In recent years, many studies have been dedicated to mitigate the effect of non-stationary acoustic noise 
in different domains \cite{ummse,tavares2016,emdh}. 
Particularly, speech enhancement solutions have been applied in the Hilbert-Huang Transform (HHT) domain.
These techniques adopt the Empirical Mode Decomposition (EMD) \cite{huang} or one of its variations 
to analyze the noisy speech signal.
This powerful decomposition has also become interesting for processing and analyze other signals, e.g. electroencephalogram signals \cite{baj19}, and multimodal sensing data \cite{sha19}.
HHT-based approaches have achieved interesting speech quality improvement in noisy scenarios 
\cite{emdh,zao_15,emdf}.
Impulsive noises may be considered as a different kind of non-stationary sources.

This letter introduces an HHT-domain method to enhance speech signals corrupted by impulsive acoustic noises. 
The proposed HHT-$\alpha$ solution applies the Ensemble EMD (EEMD) \cite{torres} to decompose a target noisy signal
into a series of intrinsic mode functions (IMF).
The noise components of each IMF are identified and selected based on the impulsiveness index $\alpha$ \cite{NikiasBook1995}
on a frame-by-frame basis.
The speech signal is reconstructed excluding frames that are mainly composed by noise.
In HHT-$\alpha$, no assumption is considered for speech and noise distributions.

Several experiments are conducted to examine the effectiveness of the proposed solution.
HHT-$\alpha$ is evaluated considering three quality and two intelligibility objective measures
that present high correlation with subjective listening tests.
Five real acoustic noises with different impulsiveness degrees are used to corrupt speech utterances.
Five values of signal-to-noise ratio (SNR) are considered in this work: 
-10 dB, -5 dB, 0 dB, 5 dB, and 10 dB.
Three speech enhancement techniques are adopted as baseline: 
the spectral Wiener filtering with unbiased minimum mean-square error estimator (UMMSE) \cite{ummse}, and 
the time domain EMD-based filtering (EMDF) \cite{emdf} and EMD-Hurst-based (EMDH) \cite{emdh} approaches.
Experiments demonstrate that the HHT-$\alpha$ method achieves interesting speech quality results, especially 
for highly impulsive noises.
HHT-$\alpha$ also shows similar average intelligibility rate when compared to the competitive techniques.

\section{HHT-$\alpha$: Speech Enhancement Scheme}

\begin{figure}[t!]
\centering
\includegraphics[width=0.8\columnwidth]{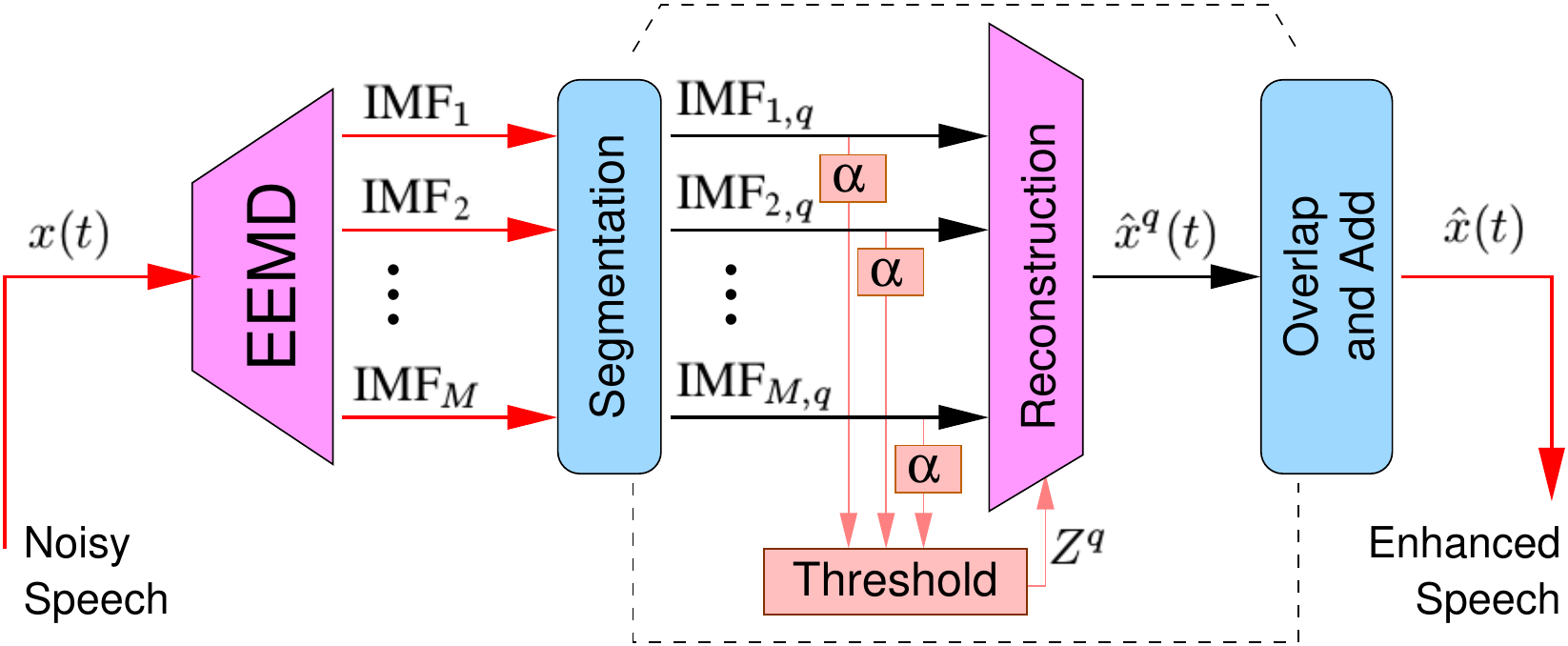}
\vspace{-0.cm}
\caption{Block diagram of the HHT-$\alpha$ speech enhancement method.}
\vspace{-0.cm}
\label{fig:diagram}
\end{figure}

The HHT-$\alpha$ speech enhancement includes three main steps:
noisy signal decomposition, estimation and selection of noise components, and speech signal reconstruction.
Fig. \ref{fig:diagram} illustrates the block diagram of the proposed method.

\subsection{Noisy Signal Decomposition}

HHT \cite{huang} is a nonlinear adaptive approach that locally analyzes a signal $x(t)$ to define a 
local high-frequency part, also called detail $d(t)$, 
and a local trend $a(t)$, such that $x(t)=d(t)+a(t)$. 
An oscillatory IMF is derived from the detail function $d(t)$.
The high versus low-frequency separation procedure is iteratively repeated over the residual
$a(t)$, leading to a new detail and a new residual. 
Thus, the decomposition leads to a series of IMFs and a residual, such that
$x(t)=\sum_{m=1}^M\mbox{IMF}_m(t)+r(t)$,
where $\mbox{IMF}_m(t)$ is the $m$-th mode of $x(t)$ and $r(t)$ is the residual.
As opposed to other kinds of signal decomposition, a set of basis functions is not demanded for the HHT.
In fact, HHT results in fully data-driven decomposition modes and 
does not require the stationarity of the target signal.

The EEMD was introduced in \cite{torres} to overcome the 
mode mixing
problem that generally occurs in the original EMD.
The key idea is to average IMFs obtained after corrupting the original signal using several realizations of 
white Gaussian noise (WGN). Thus, EEMD algorithm can be described as:
\begin{enumerate}
\item Generate $x^n(t) = x(t) + w^n(t)$, where $w^n(t)$, $n =
1, \ldots , N$, are different realizations of WGN;
\item Apply EMD to decompose $x^n(t)$, $n = 1, \ldots , N$, 
into a series of components $\mbox{IMF}_m^n(t)$, $m = 1, \ldots , M$;
\item Assign the $m$-th mode of $x(t)$ as
	$\mbox{IMF}_m(t) =\frac{1}{N}\sum_{n=1}^N \mbox{IMF}_m^n(t)\,;$
\item Finally, $x(t)=\sum_{m=1}^M\mbox{IMF}_m(t)+r(t)$,
where $r(t)$ is the residual.
\end{enumerate}

\subsection{Estimation and Selection of Noise Components}

In the literature, impulsive signals and noises are generally defined by a sequence of random samples with 
symmetric heavy-tail distribution, i.e., $P[X>x]\sim C|x|^{-\alpha}$, where $C$ is a positive constant 
and $0<\alpha\leq 2$ is the impulsiveness index.
The $\alpha$ exponent is also related to $\alpha$-stable distribution 
and may be described as
the characteristic exponent \cite{NikiasBook1995}.

In \cite{Komaty2015} authors showed that for $\alpha$-stable noises the EMD 
behaves like a quase-dyadic filterbank for $\alpha \in\, ]1.2,2.0]$.
Speech signals investigated in this work are impulsive and present heavy-tails with
$\alpha$ values in the range $[0.9, 1.2]$. On the other hand, acoustic noises commonly encountered 
in real urban scenarios have values in the range $[1.2,2.0]$ \cite{Komaty2015}.
Thus, in this letter the EMD is applied to highlight the noise impulsiveness of the corrupted speech signal.
The estimator proposed by McCulloch in \cite{McCulloch1996,McCulloch1998} is here adopted 
for the $\alpha$ index estimation.

\begin{figure}[t]
\centering
\vspace{-0.2cm}
\includegraphics[width=0.9\columnwidth,height=0.21\textheight]{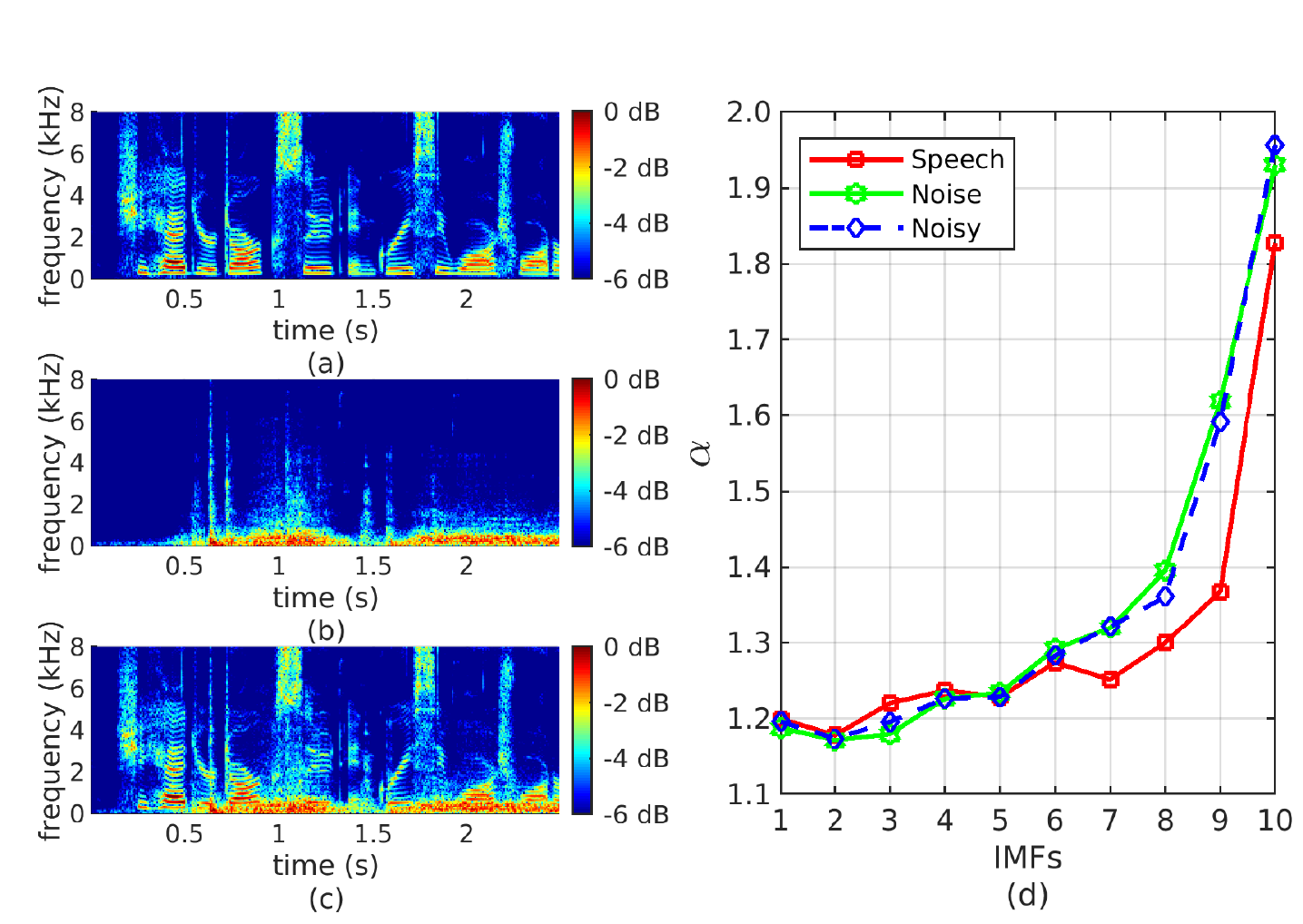}\\
\caption{Spectrograms of (a) clean speech (b) Sliding Door Closing noise ($\alpha=1.21$), and (c) noisy speech (SNR=$0$ dB). 
(d) The average values of $\alpha$ estimated from the IMFs.}
\vspace{-0.3cm}
\label{fig:selection}
\end{figure}

Fig.~\ref{fig:selection}(a)-(c) show spectrograms of a clean speech signal collected from the TIMIT database \cite{timit}, 
an impulsive Sliding Door Closing noise with $\alpha=1.21$, and also the corrupted
signal with $\mbox{SNR}=0$ dB. Note from Fig.~\ref{fig:selection}(b) that the noise energy 
is mostly concentrated at low frequencies
and the spectrogram has sharp wide band components around $0.6$, $1.0$ and $1.5$ seconds.
Fig.~\ref{fig:selection}(d) presents average values of the impulsiveness index $\alpha$
estimated from IMFs of clean speech, impulsive noise, and noisy speech signals.
It can be seen that as the mode index increases, the $\alpha$ values of all signals approach $2$.
For the highest IMF indexes, e.g., $7-10$, 
the acoustic noise and the noisy speech signal have similar $\alpha$ values.
These values are greater than those obtained from the clean speech signal.
This indicates that these IMFs are more noise-like,
which corroborates with previous works (for example, refer to \cite{emdh}).

\begin{figure}[t]
\begin{center}
\hspace{-.2cm}
\includegraphics[width=0.265\columnwidth]{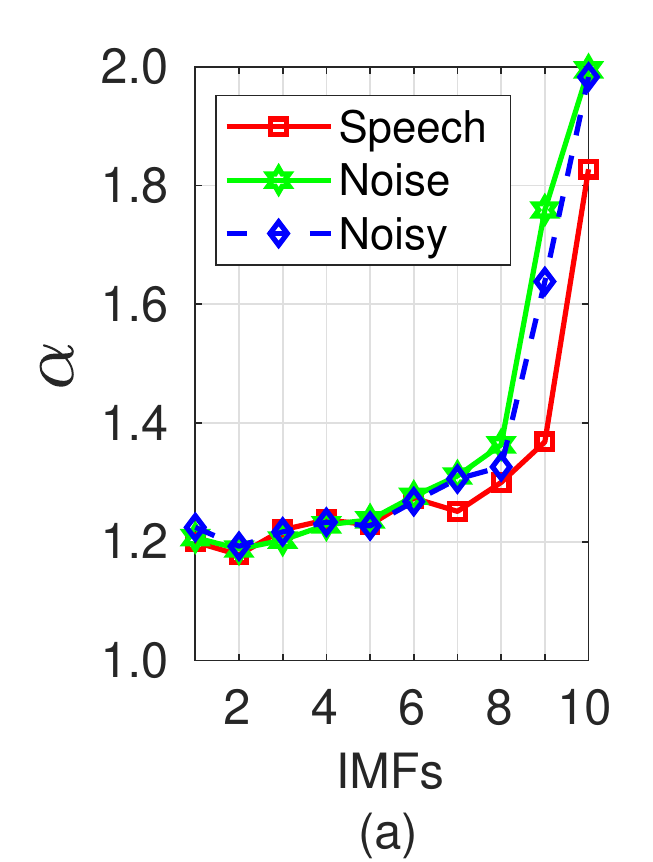}\hspace{-.2cm}
\includegraphics[width=0.265\columnwidth]{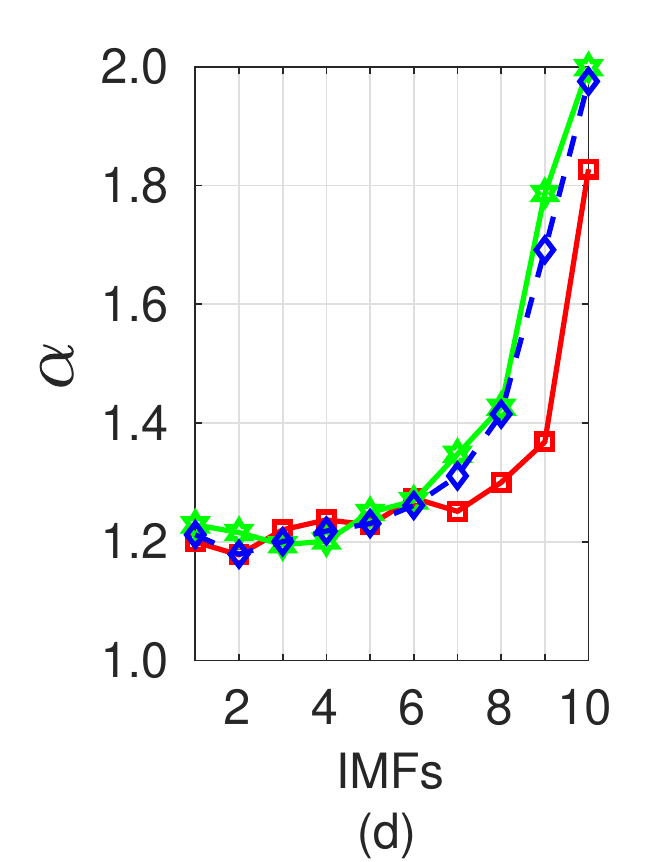}\hspace{-.2cm}
\includegraphics[width=0.265\columnwidth]{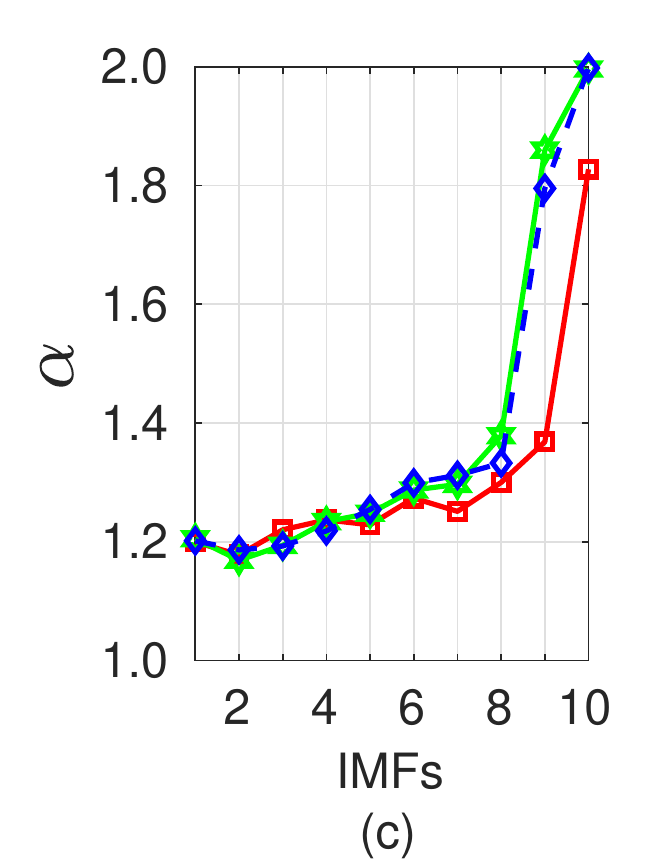}\hspace{-.2cm}
\includegraphics[width=0.265\columnwidth]{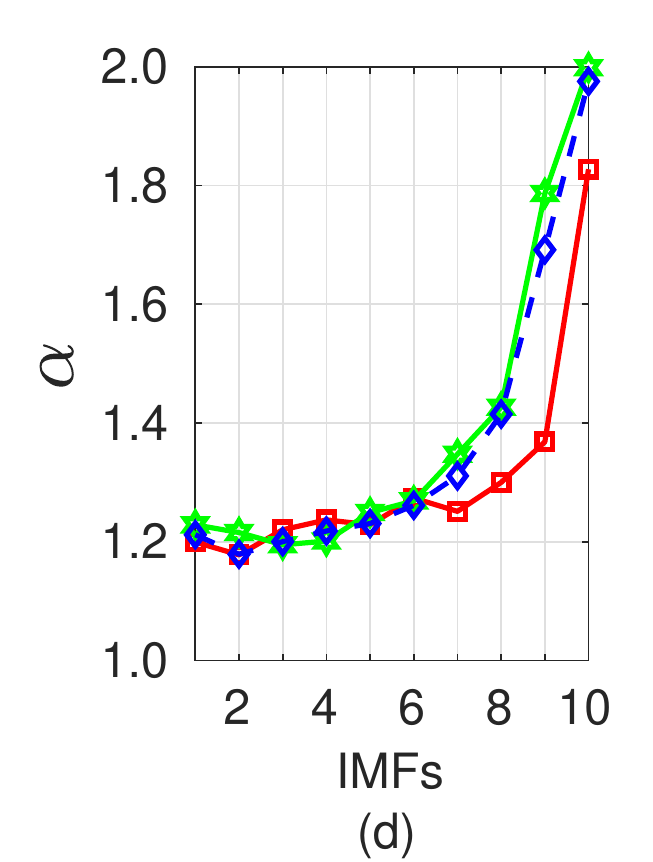}\hspace{-.2cm}\\
\caption{Average values of $\alpha$ estimated from the IMFs of impulsive noise sources:
(a) Train, (b) Horn, (c) Babble, and (d) Helicopter.}
\vspace{-0.5cm}
\label{fig:alpha_noises}
\end{center}
\end{figure}

Similar behavior can be observed in Fig. \ref{fig:alpha_noises}, where the estimated values of $\alpha$ from 
different IMFs are shown for the other four impulsive noises: Train ($\alpha=1.46$), Horn ($\alpha=1.59$), 
Babble ($\alpha=1.79$), and Helicopter ($\alpha=1.98$). Once again, $\alpha$ values indicate
that IMFs with high indices are mostly composed by noise.
Note from Fig.~\ref{fig:selection}(d) and Fig. \ref{fig:alpha_noises} that 
for medium IMF indexes, i.e., $3-5$, 
$\alpha$ values of the noisy signal generally vary between those estimated from the 
noise and from the clean speech signal. 
This demonstrates that the impulsiveness index is an appropriate identification criterion to 
select the IMFs with more speech-like characteristics and reject the noise-like components.

\begin{figure*}[!t]
\begin{center}
\includegraphics[width=2\columnwidth]{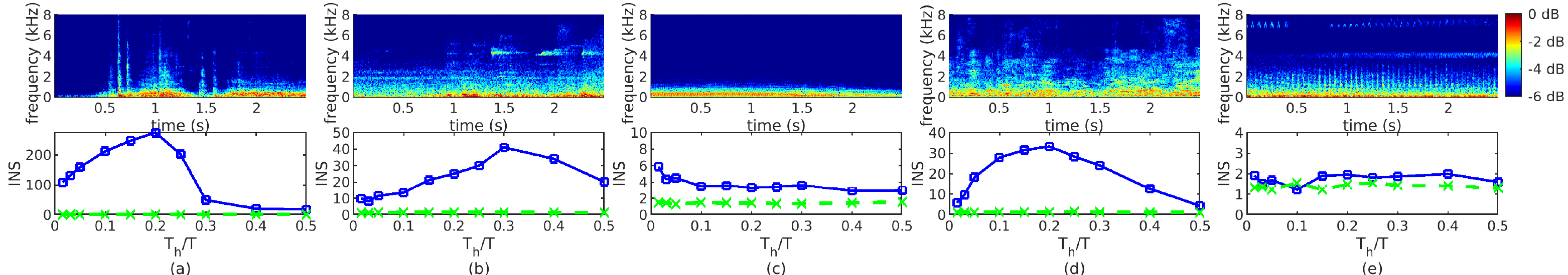}
\caption{Spectrograms and INS obtained for 2.4-seconds segments of the acoustic impulsive noises:
(a) Sliding Door Closing, (b) Train, (c) Horn (d) Babble and (e) Helicopter. Dashed lines indicate the
value for the stationarity test threshold.}
\vspace{-0.2cm}
\label{fig:ins}
\end{center}
\end{figure*}

The selection of noise components is performed as follows. 
After the decomposition of the target noisy signal with the EEMD algorithm, 
each mode $\mbox{IMF}_m$ is segmented into a set of $Q$ overlapping short-time 
frames $\mbox{IMF}_{m,q}$,
$q\in\{1, \ldots , Q\}$, with $T_d$ samples each.
In this proposal, the selection of noisy components is based on $\alpha$ parameters of each windowed IMF. 
For each frame $q$, the impulsiveness index is estimated from the decomposition modes $\mbox{IMF}_{m,q}(t)$ 
leading to a set of values $\alpha_1^q , \ldots , \alpha_M^q$. 
The next step is to determine the index $Z^q$ of the last
IMF whose impulsiveness index is bellow a given threshold, $\rho_\alpha$, i.e., $\alpha_Z^q\leq \rho_\alpha$. 
IMFs whose $\alpha$ values exceed the threshold are considered as noise-like components.

\subsection{Speech Signal Reconstruction}
\label{sec:proposed}

If $\hat{x}(t)$ represents the enhanced speech signal,
then each frame is reconstructed by
$\hat{x}^q(t)=\sum_{m=1}^{Z^q}w(t) \, \mbox{IMF}_{m,q}(t), q=1, \ldots, Q,$
where $Z^q$ is the index of the last mode considered as speech and $w(t)$ is a window function used 
to avoid discontinuities in the reconstructed signal (for more details see \cite{emdh}).
Finally, $\hat{x}(t)$ is reconstructed by overlapping and adding 
all frames as
	$\hat{x}(t)=\frac{1}{P}\sum_{q=1}^Q\hat{x}^q(t-qS_d) \, ,$
where $P$ is a normalization factor that depends on the window function $w(t)$, the frame length $T_d$,
and the step size $S_d$.

\section{Evaluation Experiments}

Extensive speech enhancement experiments are conducted
with a subset of $183$ speech segments of the
TIMIT speech database \cite{timit}.
Speech utterances have sampling rate of $16$ kHz and average time duration of $2.4$ seconds. 
Five impulsive non-stationary acoustic noises are used to corrupt the speech utterances:
Sliding Door Closing, Train, Horn, 
and Helicopter are selected from Freesound.org\footnote{Available at https://freesound.org.}, while 
Babble is obtained from the RSG-10 \cite{rsg} database.
These files are also available at lasp.ime.eb.br.

Fig.~\ref{fig:ins} presents the spectrogram and the index of non-stationarity (INS) \cite{flan10}
obtained from segments of five acoustic noises. The INS value is here shown to
objectively examine the non-stationarity of impulsive noises.
The time scale $T_h/T$ is the ratio of the length of the short-time
spectral analysis ($T_h$) and the total time duration ($T=2.4$ seconds) 
of noise sample sequences. 
For each window length $T_h$, a threshold is defined to guarantee the 
stationarity assumption with a confidence degree of $95$\%. 
Thus, if $\mbox{INS}\leq \gamma$ then the noise is considered as stationary. Otherwise, it is designated as non-stationary.
The $\gamma$ values are also exhibited in Fig.~\ref{fig:ins}.

Sliding Door Closing, Train, and Babble noises 
are here classified as highly non-stationary since
their INS achieves values greater than $200$, $40$, and $30$, respectively.
Horn noise presents INS results in the range $\left[3, 6\right]$ and thus,
it is defined as moderately non-stationary. 
Helicopter noise is considered as stationary since the INS values are quite similar to the stationarity threshold
for all time scales.

The performance of the proposed and baseline methods are examined using five objective measures.
Perceptual evaluation of speech quality (PESQ), log-likelihood ratio (LLR), and frequency-weighted segmental SNR (fwSNRseg) \cite{hu2008} are used to evaluate enhanced speech signals in terms of quality.
These measures present high correlation with subjective overall quality and
signal distortion results \cite{hu2008}.
Coherence speech intelligibility index (CSII) \cite{csii} and 
short-time objective intelligibility measure (STOI) \cite{stoi}
are adopted for speech intelligibility assessment. 
Intelligibility prediction scores are obtained according to the mapping function
$f(d)=\frac{100}{1+\exp{a \,d + b}}$, where $d$ refers to the objective measure.
In this work, it is adopted $a=-10.09$ and $b=4.65$ for the CSII, and $a=-13.45$ and $b=9.36$ for the STOI.

For the HHT-$\alpha$ method, the EEMD algorithm is applied considering 50 different realizations of WGN
with SNR of 30 dB to obtain 10 IMFs.
The decision threshold $\rho_\alpha$ is crucial to determine the components to be removed 
from each corrupted speech frame. In this letter, an adaptive threshold is introduced, such that
$\rho_\alpha = \min (\mu \alpha^q_u, \alpha_{\text{min}})$, where $\mu=0.8$, $\alpha^q_u$ is the estimate of $\alpha$ 
for the corrupted speech windowed signal, and $\alpha_{\text{min}}$ is the minimum value 
allowed for $\rho_\alpha$. In this work, $\mu$ is adopted to adjust the amount
of noise components to be removed, while $\alpha_{\text{min}} = 1.1$ is used to avoid excessive component removal 
in speech dominant segments of the signal.
The selection of noise components considers $T_d=10240$ samples per frame
and step size of $S_d=128$ samples.

Tab.~\ref{tab:pesq} shows the PESQ results obtained with the proposed and 
baseline speech enhancement techniques for different impulsive acoustic noises and SNR values. 
Note that HHT-$\alpha$ outperforms the competing time domain approaches for most of the noisy scenarios. Particularly for the Sliding Door Closing noise, which presents the
lowest $\alpha$ value, the HHT-$\alpha$ achieves the highest average PESQ result, including the spectral 
UMMSE method.
On average, the overall PESQ obtained with the proposed solution is 1.93, which is 0.04 and 0.10 higher than 
EMDH and EMDF, respectively.
The spectral UMMSE achieved an overall PESQ of 2.11.

{
\begin{table}[t]
\renewcommand{\arraystretch}{1.}
\begin{center}
\caption{PESQ results with the proposed and baseline methods.}
\vspace{-0.2cm}
{\scriptsize
\begin{tabular}{|c|r|c|c|c|c|} \hline
{Noise} & \multicolumn{1}{c|}{SNR} &{UMMSE} & {EMDF} & {EMDH} & {HHT-$\alpha$}\\ \hline \hline
\multirow{5}{*}{\begin{tabular}{c}{Sliding Door}\\{Closing}\vspace{0.1cm}\\$\alpha = 1.21$\end{tabular}}
	& $-10$ & $1.19$ & $1.17$ & $\bf 1.21$ & $1.20$ \\ \cline{2-6}
	& $-5$ & $1.58$ & $1.51$ & $1.58$ & $1.58$ \\ \cline{2-6}
	& $0$ & $2.00$ & $1.90$ & $1.98$ & ${\bf 2.03}$ \\ \cline{2-6}
	& $5$ & $2.37$ & $2.22$ & $2.35$ & ${\bf 2.44}$ \\ \cline{2-6}
	& $10$ & $2.69$ & $2.53$ & $2.68$ & ${\bf 2.78}$ \\ \hline
\multirow{5}{*}{\begin{tabular}{c}{Train}\vspace{0.1cm}\\$\alpha = 1.46$\end{tabular}}
	& $-10$ & ${\bf 1.21}$ & $1.04$ & $1.08$ & $0.99$ \\ \cline{2-6}
	& $-5$ & ${\bf 1.74}$ & $1.47$ & $1.50$ & $1.48$ \\ \cline{2-6}
	& $0$ & ${\bf 2.18}$ & $1.90$ & $1.92$ & $1.93$ \\ \cline{2-6}
	& $5$ & ${\bf 2.56}$ & $2.29$ & $2.31$ & $2.34$ \\ \cline{2-6}
	& $10$ & ${\bf 2.89}$ & $2.61$ & $2.67$ & $2.67$ \\ \hline	
\multirow{5}{*}{\begin{tabular}{c}{Horn}\vspace{0.1cm}\\$\alpha = 1.59$\end{tabular}}
	& $-10$ & ${\bf 1.75}$ & $1.35$ & $1.42$ & $1.44$ \\ \cline{2-6}
	& $-5$ & ${\bf 2.12}$ & $1.61$ & $1.70$ & $1.84$ \\ \cline{2-6}
	& $0$ & $\bf {2.46}$ & $1.95$ & $2.05$ & $2.23$ \\ \cline{2-6}
	& $5$ & ${\bf 2.78}$ & $2.25$ & $2.38$ & $2.59$ \\ \cline{2-6}
	& $10$ & ${\bf 3.09}$ & $2.56$ & $2.69$ & $2.85$ \\ \hline	
\multirow{5}{*}{\begin{tabular}{c}{Babble}\vspace{0.1cm}\\$\alpha = 1.79$\end{tabular}}
	& $-10$ & $0.91$ & $0.91$ & $0.91$ & $0.86$ \\ \cline{2-6}
	& $-5$ & $\bf 1.33$ & $1.24$ & $1.25$ & $1.22$ \\ \cline{2-6}
	& $0$ & ${\bf 1.76}$ & $1.58$ & $1.60$ & $1.61$ \\ \cline{2-6}
	& $5$ & ${\bf 2.17}$ & $1.94$ & $1.99$ & $1.97$ \\ \cline{2-6}
	& $10$ & ${\bf 2.54}$ & $2.27$ & $2.35$ & $2.28$ \\ \hline	
\multirow{5}{*}{\begin{tabular}{c}{Helicopter}\vspace{0.1cm}\\$\alpha = 1.98$\end{tabular}}
	& $-10$ & ${\bf 1.51}$ & $1.17$ & $1.18$ & $1.26$ \\ \cline{2-6}
	& $-5$ & ${\bf 1.92}$ & $1.52$ & $1.54$ & $1.62$ \\ \cline{2-6}

	& $0$ & ${\bf 2.30}$ & $1.91$ & $1.93$ & $1.99$ \\ \cline{2-6}
	& $5$ & ${\bf 2.66}$ & $2.29$ & $2.31$ & $2.35$ \\ \cline{2-6}
	& $10$ & ${\bf 2.99}$ & $2.61$ & $2.66$ & $2.65$ \\ \hline	
\end{tabular}}
\label{tab:pesq}
\end{center}
\vspace{-.5cm}
\end{table}
}

\begin{figure}[t!]
\centering
\includegraphics[width=0.75\columnwidth]{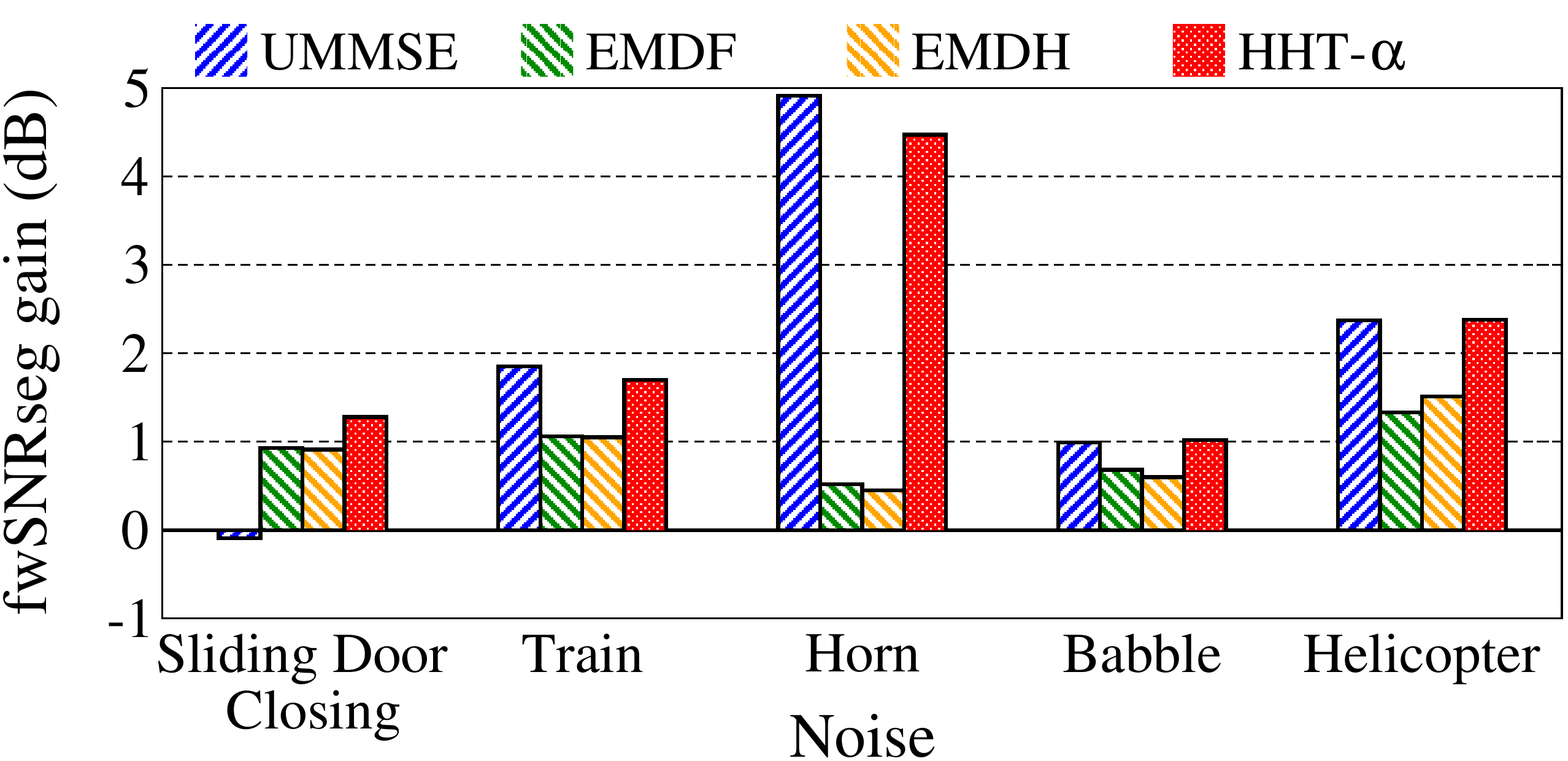}
\vspace{-0.1cm}
\caption{Average fwSNRseg gain obtained for different noise sources.}
\label{fig:fwsnrseg}
\vspace{-0.3cm}
\end{figure}

Fig. \ref{fig:fwsnrseg} exhibits the average fwSNRseg improvement obtained by the proposed and baseline methods
for the five noises.
Once again, HHT-$\alpha$ achieves the best results for the highly impulsive Sliding Door Closing noise.
It is interesting to mention that, for this noise, 
UMMSE does not improve the speech signals in terms of fwSNRseg.
For the other noise sources, HHT-$\alpha$ outperforms the time domain EMDF and EMDH techniques.
Moreover, the fwSNRseg gain of HHT-$\alpha$ is slightly superior than that obtained with UMMSE for the Babble and 
Helicopter noises.

\begin{figure}[t!]
\centering
\includegraphics[width=0.75\columnwidth]{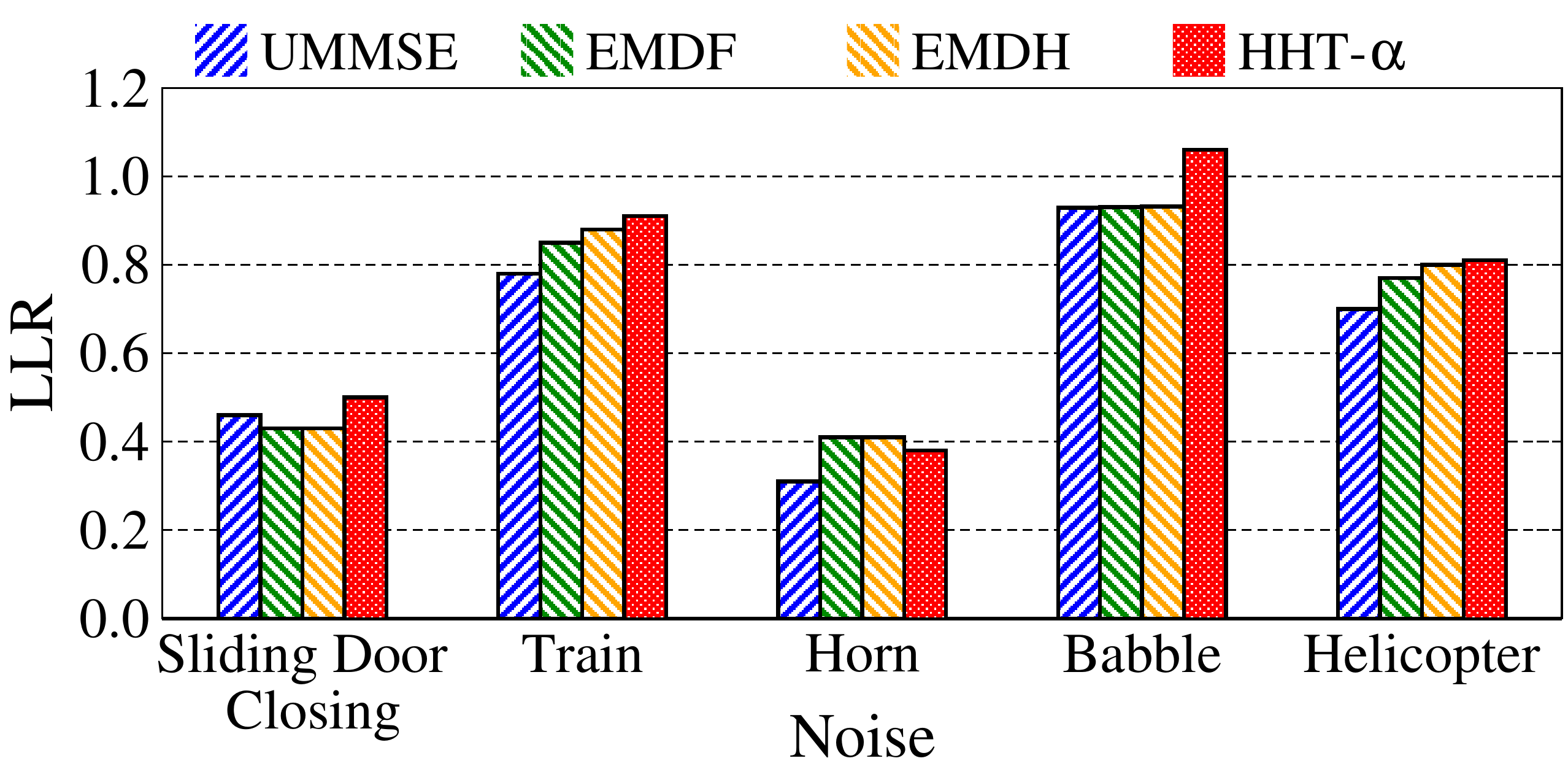}
\vspace{-0.1cm}
\caption{Average LLR results obtained for different noise sources.}
\vspace{-0.1cm}
\label{fig:llr}
\end{figure}

{
\begin{table}[t!]
\renewcommand{\arraystretch}{1.}
\begin{center}
\caption{STOI Intelligibility rate prediction (\%).}
\vspace{-0.2cm}
{\scriptsize
\begin{tabular}{|c|r|c|c|c|c|} \hline
{Noise} & \multicolumn{1}{c|}{SNR} &{UMMSE} & {EMDF} & {EMDH} & {HHT-$\alpha$}\\ \hline \hline
\multirow{5}{*}{\begin{tabular}{c}{Sliding Door}\\{Closing}\vspace{0.1cm}\\$\alpha = 1.21$\end{tabular}}
& $-10$ &  $15.6 $ & $\bf 16.7 $ & $ 16.5 $ & $ 15.0 $ \\ \cline{2-6}
& $-5$ &  $35.4 $ & $\bf 37.2$ & $36.7$ & $34.8$ \\ \cline{2-6}
& $0$ &  $59.3$ & $\bf 60.8$ & $60.5$ & $60.1$ \\ \cline{2-6}
& $5$ &  $77.8$ & $78.6$ & $78.3$ & $\bf 78.9 $ \\ \cline{2-6}
& $10$ &  $88.6$ & $\bf 88.9$ & $88.7$ & $88.6 $ \\ \hline
\multirow{5}{*}{\begin{tabular}{c}{Train}\vspace{0.1cm}\\$\alpha = 1.46$\end{tabular}}
& $-10$ &  $12.2$ & $12.8$ & $\bf 13.8$ & $10.1 $ \\ \cline{2-6}
& $-5$ &  $32.8$ & $33.8$ & $\bf 34.4$ & $28.9 $ \\ \cline{2-6}
& $0$ &  $60.6$ & $\bf 61.9$ & $61.8$ & $57.2 $ \\ \cline{2-6}
& $5$ &  $81.1$ & $\bf 82.1$ & $81.5$ & $78.9 $ \\ \cline{2-6}
& $10$ &  $91.3$ & $\bf 91.4$ & $91.2$ & $89.7 $ \\ \hline
\multirow{5}{*}{\begin{tabular}{c}{Horn}\vspace{0.1cm}\\$\alpha = 1.59$\end{tabular}}
& $-10$ &  $47.6$ & $\bf 53.6$ & $53.4$ & $37.0 $ \\ \cline{2-6}
& $-5$ &  $60.6$ & $\bf 65.1$ & $65.0$ & $55.8 $ \\ \cline{2-6}
& $0$ &  $72.1$ & $\bf 74.8$ & $74.5$ & $71.9 $ \\ \cline{2-6}
& $5$ &  $82.3$ & $\bf 83.1$ & $82.7$ & $82.6 $ \\ \cline{2-6}
& $10$ &  $\bf 90.4$ & $89.9$ & $89.8$ & $89.1 $ \\ \hline
\multirow{5}{*}{\begin{tabular}{c}{Babble}\vspace{0.1cm}\\$\alpha = 1.79$\end{tabular}}
& $-10$ &  $3.0$ & $4.9$ & $4.9$ & $3.8 $ \\ \cline{2-6}
& $-5$ &  $11.1$ & $\bf 15.1$ & $14.9$ & $14.8 $ \\ \cline{2-6}
& $0$ &  $35.4$ & $\bf 40.1$ & $39.4$ & $39.7 $ \\ \cline{2-6}
& $5$ &  $69.3$ & $\bf 70.9$ & $70.2$ & $68.5 $ \\ \cline{2-6}
& $10$ &  $\bf 88.4$ & $88.2$ & $88.0$ & $85.3 $ \\ \hline
\multirow{5}{*}{\begin{tabular}{c}{Helicopter}\vspace{0.1cm}\\$\alpha = 1.98$\end{tabular}}
& $-10$ & $\bf 20.4$ & $15.0$ & $16.8$ & $16.7 $ \\ \cline{2-6}
& $-5$ &  $\bf 45.6$ & $37.7$ & $39.1$ & $41.8 $ \\ \cline{2-6}
& $0$ &  $\bf 72.1$ & $66.0$ & $66.0$ & $67.6 $ \\ \cline{2-6}
& $5$ &  $\bf 87.9$ & $84.6$ & $84.5$ & $83.3 $ \\ \cline{2-6}
& $10$ &  $\bf 94.4$ & $93.2$ & $93.1$ & $91.4 $ \\ \hline 
\end{tabular}}
\label{tab:stoi}
\end{center}
\vspace{-.2cm}
\end{table}
}

\begin{figure}[t!]
\begin{center}
\includegraphics[width=\columnwidth]{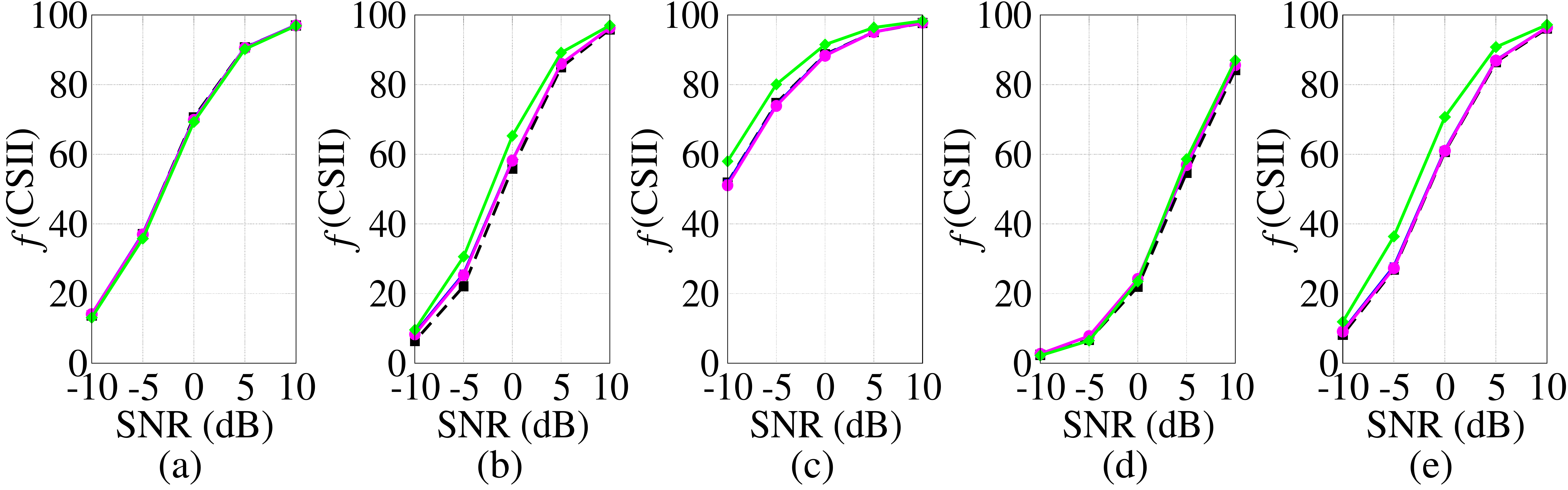}\\
\vspace{-0.2cm}
\caption{CSII intelligibility prediction rates obtained for (a) Sliding Door Closing,
(b) Train, (c) Horn, (d) Babble, and (e) Helicopter acoustic noises.}
\vspace{-0.3cm}
\label{fig:csii}
\end{center}
\end{figure}

Fig. \ref{fig:llr} depicts the average LLR values obtained for each impulsive noise.
Note that the proposed solution again achieves the highest LLR for four noise sources.
The only exception is the Horn noise. However, HHT-$\alpha$ outperforms the spectral UMMSE for this noise source.
The overall LLR obtained with HHT-$\alpha$ is 0.73, which is 0.04, 0.05 and 0.09 higher than results achieved with 
EMDH, EMDF and UMMSE, respectively.

Tab.~\ref{tab:stoi} presents intelligibility prediction rates obtained with STOI.
Note that HHT-$\alpha$ and competitive solutions achieve quite close results,
especially for SNR $\geq 0$ dB.
On average, intelligibility prediction rates vary in at most 2.2 percentage points, 
i.e., from 57.0\% with HHT-$\alpha$ to 59.2\% with EMDH. 
This similar behavior in terms of speech intelligibility is reinforced by CSII results depicted in Fig.~\ref{fig:csii}.
Once again, the proposed and baseline algorithms show similar speech intelligibility prediction values.

\vspace{-.2cm}
\section{Conclusion}
This letter introduced the HHT-$\alpha$ speech enhancement technique based on the Hilbert-Huang Transform.
The EEMD algorithm is used to decompose the noisy speech signal in time domain.
The estimation and selection of noise components is performed frame-by-frame 
based on the impulsiveness index of the decomposition modes.
The enhanced version of the speech signal is finally reconstructed using the IMFs
that are mainly composed of speech.
Several experiments were conducted using five non-stationary acoustic noises with different 
values of the impulsiveness index $\alpha$.
Particularly for the most impulsive noise, the proposed solution outperformed the three competing approaches
in terms of PESQ, fwSNRseg, and LLR objective quality measures.
In terms of speech intelligibility, HHT-$\alpha$ is similar to other state-of-the-art methods 
for all the impulsive noise sources.

\vspace{-0.2cm}
\section*{Acknowledgment}
\scriptsize
\addcontentsline{toc}{section}{Acknowledgment}
R. Coelho is partially supported by the 
National Council for Scientific and Technological Development (CNPq) 307866/2015 and
Fundação de Amparo à Pesquisa do Estado do Rio de Janeiro (FAPERJ) 203075/2016
research grants.

\normalsize

\vspace{-.2cm}
\bibliographystyle{IEEEtran}
\bibliography{IEEEabrv,refs}

\end{document}